# Optimizing spin polarization in quantum dot vertical-gain structures through pump wavelength selection

Najm Alhosiny[1]*, Sami S. Alharthi[1], J. Doogan[2,3,5], E. Clarke[4], and T. Ackemann[2]

[1] *Department of Physics, College of Science, Taif University, P.O. Box 11099, Taif 21944, Saudi Arabia*
[2] *SUPA and Department of Physics, University of Strathclyde, Glasgow G4 0NG, Scotland, United Kingdom*
[3] *M Squared Lasers Ltd., Glasgow G20 0SP, UK*
[4] *EPSRC National Epitaxy Facility, University of Sheffield, Sheffield S3 7HQ, UK*
[5] *now at: Trumpf Laser + Machinery Ireland Limited, Dublin, D24 W702, Ireland*

* Corresponding author: Najm Alhosiny (najm@tu.edu.sa)

*Spintronic applications require an efficient injection of spin-polarized carriers. We study the maximally achievable spin polarization in InAs quantum dots in a vertical-cavity gain structure to be used in telecoms-wavelength vertical-external-cavity surface-emitting lasers via measurement of the Stokes parameter of the photoluminescence emission around 1290 nm. Using five pump wavelengths between 850 and 1070 nm, the observed spin polarization depends strongly on the pump wavelength with the highest polarization of nearly 5% found for excitation at 980 nm. This corresponds to an effective spin lifetime of 40 ps and is attributed to the dominant excitation of heavy holes only.*

Over the last few decades, there has been considerable interest in spin-injection into semiconductor lasers with vertical-cavity gain structures, in particular Vertical-Cavity Surface-Emitting LASERs (VCSELs) [1-2]. Due to optical selection rules, spin-VCSELs offer the intriguing possibility of controlling the polarization state of the light output via the spin-polarization of the injected carriers [3-9] and related characteristics including lasing threshold reduction [10-11], spin amplification [12] and high-frequency oscillatory behavior without the need for an external perturbation or drive [1, 13-15]. Polarization modulation in spin lasers enables modulation faster than the relaxation oscillation frequency and hence has promise for future ultrafast communication systems [1, 16-17]. These intriguing features make spin-VCSEL a promising candidate as a building block in information technology and applications utilizing coherent light-matter interactions. Among the recent proposed applications are frequency comb generation [18], photonic signal generation [19], local oscillators for homodyne detection [20], and reservoir computing [21-23].

From a practical point of view, the efficient generation, storage and detection of spin-polarized carriers is still a challenge, in particular the ultimate goal of spin-VCSELs achieving room temperature (RT) operation with pure electrical spin injection [24, 25]. Various approaches have been employed for spin injection into these devices [26], ranging from ferromagnetic contacts [5], chiral metastructures [27] to tunnel injectors [28], but for many purposes creating spin polarization via spin polarized optical pumping utilizing the selection rules is still the preferable method. Efficient injection of spin-polarized carriers into Quantum Well (QW) [3, 4, 6, 8] and Quantum Dot (QD) [7] spin-V(E)CSELs with optical polarization up to 100% after spin amplification under lasing conditions has been demonstrated. QW-based spin-VCSELs have been widely explored and they represent the majority of the demonstrated spin-VCSEL devices [6, 8, 9, 17, 29-32]. However, their QD-based counterparts possess promise for spin-devices as the D'yakonov–Perel spin-relaxation mechanism limiting spin polarization in quantum wells is suppressed in QD [24, 25, 33-34]. Spin lifetime and dynamics have been examined in undoped [33, 35-37], p-doped [36, 37-40], and n-doped [37, 41-44] InAs/GaAs QDs. At low temperature, often in conjunction with a magnetic field, very long spin relaxation time from ns to a record of 20 ms can be obtained [33, 35, 39, 45-47]. Within working VCSEL structures, Basu *et al.* reported spin relaxation times of 100 - 150 ps in InAs/GaAs QD spin-VCSELs under magnetic fields of a few Tesla at 200 K [48]. At room temperature, reported lifetimes are 70-80 ps [26, 47, 49] (up to 120 ps [38]) for InAs/GaAs QDs, but not constituting dots-in-a-well (DWELL) samples. However, most investigations on QD structures either demonstrate electrical spin injection concepts or are performed on single layers of low density QDs, suitable to isolate single QDs for quantum technology applications but not suitable as laser gain structures.

By studying the polarization resolved photoluminescence (PL) while exciting the samples at different pump wavelengths, we investigate excitation conditions for high spin polarization in InAs/GaAs QDs incorporated into a gain structure for vertical-external cavity surface-emitting lasers (VECSELs). These structures achieve about 80% spin polarization via spin amplification under lasing condition in a VECSEL [7]. VECSELs are a versatile extension of the VCSEL design where the external free space cavity allows for the insertion of intracavity optical elements such as semiconductor saturable absorber mirrors (SESAMs), polarization control and nonlinear crystals. This flexible configuration can be customized to allow for a wide range of functionalities and power scaling [50]. Spin-optoelectronics has focused typically on VCSELs but there is interest in other micro-VECSELs where the external reflector is a fiber end and the cavity length on the order of micrometers [7] and "macroscopic" VECSELs or semiconductor disk lasers similar in design to optically pumped solid-state lasers [6, 51, 52]. VECSELs are typically pumped optically for injection homogeneity so optical spin injection is readily achievable and optimized excitation conditions can be investigated.

Figure 1 (a) shows a schematic diagram of a QD-based vertical-cavity gain structure for a VECSEL (see [7, supplementary material] for details). The sample was grown by molecular beam epitaxy on an undoped GaAs(100) substrate, comprising a Distributed Bragg Reflector (DBR) with 25 alternating layers of AlAs and GaAs providing high reflection between 1240 and 1370 nm, a GaAs cavity containing five groups of three QD layers, each positioned around an antinode of the cavity providing resonant periodic gain. Each QD layer consists of InAs QDs embedded in a 7 nm $In_{0.12}Ga_{0.88}As$ QW DWELL structure [53]. Barriers and spacers between the DWELL layers consist of GaAs. The InAs QD density, as measured by atomic force microscopy of an uncapped QD layer grown under similar conditions as the layers in this structure, was $4\times10^{10}$ cm$^{-2}$.

Figure 1 (b) shows the complete experimental setup.

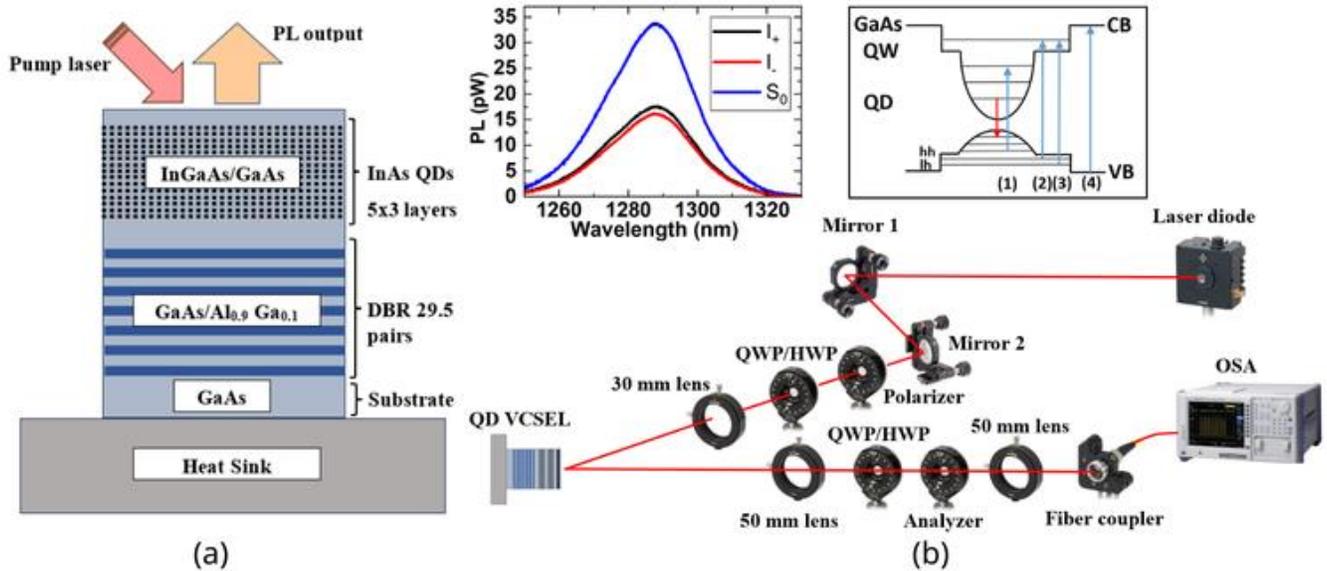

FIG. 1. (a) Schematic diagram of the half-VCSEL used in this study, (b) the experimental setup. The insets show (left): the PL of the sample (pumped with 980 nm, resolution bandwidth (RBW) 10 nm, averaged over 50 runs), (right): schematic diagram of energy states in the QD DWELL system, with excitation by the variable wavelength pump (blue arrows) into (1) QD excited state, (2) e1-hh1 transition in the QW, (3) e1-lh1 transition in the QW, or (4) bulk GaAs, and emission (red arrow) from the QD ground state.

As pump lasers, we use spatial single-mode edge-emitting lasers at 905-980 nm (Axcel M9-905-0200-D5P with 160 mW at sample, Axcel M9-915-0300-D5P, 220 mW; Thorlabs L980H1, 240 mW), an external cavity diode laser at 852 nm (Toptica DLS Pro, 100 mW), and a fiber laser at 1070 nm (IPG YLM-10-LP-SC, 1.5 W). The pump beam is aligned via a series of highly reflective mirrors. A well-defined linear polarization state is created via a Glan-Thompson polarizer. An achromatic quarter-wave plate (QWP) (Thorlabs AQWP05M-980, 690-1200 nm) follows to control the helicity of the pump beam. The beam is then focused onto the sample at an angle of incidence of 30° to the normal with a f=30 mm achromatic lens. The spot size on the sample is estimated to be 20-40 μm (radius at 1/e$^2$-point). The sample is mounted on a translatable brass plate with a thermocouple sensor to control the temperature and held at 16°C. The normal luminescence from the sample along with some scattered pump light is collimated using a f=50 mm plano-convex lens (1'' diam.) before passing through an achromatic QWP (AQWP05M-1600, 1100-2000 nm) and a Glan-laser polarizer (Thorlabs Glan-laser GL15C), which serve as the analysis tool of the experiment. The beam is then focused (using a 50 mm aspheric lens) and coupled into a multi-mode fiber to be sent to an Optical Spectrum Analyzer (OSA) (HP86140A). The spectral resolution is important not only to discriminate against weak PL from higher quantum dot states but also to remove scattered pump light which contaminates the signal although the specular reflection is not collected by the collimating lens. After optimization of fiber coupling, optical spectra are recorded on the OSA, typically averaging 10 spectra. The background noise floor of the OSA is subtracted from the measured spectra. Representative spectra are shown as inset in Fig. 1b, showing emission of the ground state of the QD ensemble. The peak PL is at 1285-1295 nm depending on the position on wafer. For pump wavelengths of 905-980 nm the detected total PL power $S_0$ in the peak is about 25-100 pW in 10 nm resolution bandwidth, depending on details of alignment, wafer, position on wafer and pump wavelength. For 852 nm, it is about 10 pW, for 1070 nm 25 pW at their respective maximal pump powers given above. Due to the desire to minimize the energy difference between pump and emission (and hence heating) and achieving a high pump efficiency, the most interesting wavelength to pump for actual laser operation are the ones pumping the quantum well (i.e 905-980 nm). 980 nm was used in Ref. [7].

In our investigations, we measure the Stokes parameters describing the polarization state of the photoluminescence and infer from this spin polarization and spin lifetimes. The experimental method detailed in Ref. [54] is used to measure Stokes parameters. We concentrate on the S3 parameter as it characterizes light helicity, respectively spin:

$$S_0 = \sigma_+ + \sigma_- , \tag{1}$$

$$S_3 = \frac{\sigma_+ - \sigma_-}{S_0} , \tag{2}$$

where $\sigma_+$ and $\sigma_-$ represent the intensity of left circularly polarized (LCP) and right circularly polarized (RCP) light, respectively. $S_0$ represents the total intensity for a given measurement. $\sigma_+, \sigma_-$ are measured by positioning a QWP with the principal axis at an angle of ±45° to the transmissive axis of the analyzer. A limitation of the measurement is that because we need to measure via the OSA, there is only one input channel and hence the measurements of $\sigma_+, \sigma_-$ are not taken simultaneously and will be susceptible to fluctuations and drifts of the PL intensity due to fluctuations of pump power and spectra. We characterized the combination of QWP and analyzer by analyzing the polarization state of a narrow-band tunable fiber-coupled laser at 1295 nm. The beam is linearly polarized with a Glan-Thompson polarizer and should have zero $S_3$. Over 20 measurements we find an average $S_3$ of -0.0005 with a standard deviation of 0.005 and a span of ±0.006. This gives an indication of the sensitivity of the experiment under best conditions.

Using the method laid out previously, the Stokes parameters are calculated. Figure 2 (a) shows an example of the calculated $S_3$ for linear and circular pump. There is a rather flat range between 1280 and 1300 nm with relatively low noise around the peak of the PL (here at 1290 nm). Outside of the 1280-1300 nm window, in the wings of the peak, noise increases as the signal-to-noise ratio decreases. Hence the average over the central 20 nm is taken as the reading and the standard deviation to characterize the uncertainty. This results in a value of $S_3 \approx -0.008\pm0.006$ for linear pump polarization, $S_3 \approx -0.039\pm0.006$ for $\sigma_-$ pump polarization, $S_3 \approx 0.035\pm0.005$ for $\sigma_+$ pump polarization. The curves for the pumps of opposite helicity are clearly separated and box in the result for linearly polarized pump, as expected. This provides strong evidence for spin memory in QDs. The small deviations from the expected symmetry around zero are due to intensity fluctuations between measurements of the circularly polarized components and the uncertainty of aligning the angle of the quarter-wave plate [supplementary material].

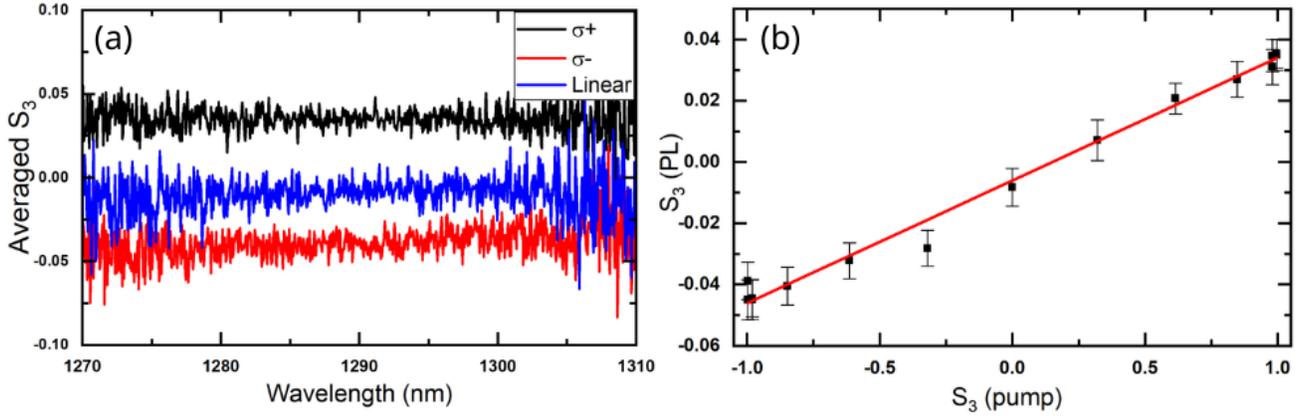

FIG. 2. (a) Example for $S_3$ obtained for linear and circular pump cases. The PL peak is at 1290 nm and the averaging was carried out in the range 1280 – 1300 nm. (b) PL $S_3$ vs pump $S_3$ and linear regression curve (red line). The pump laser is at 980 nm.

Figure 2 (b) shows an example of the relation between pump $S_3$ and PL $S_3$ when pumping with 980 nm. $S_3$ at each point is averaged over a 20 nm range and the error bar shown is the standard deviation of the distribution, i.e. represents only the statistical uncertainty. It can be seen from the figure that there is a clear linear relationship between the $S_3$ of the pump light in the sample and the $S_3$ of the PL, which again provides evidence of spin memory. Some asymmetry around the zeroes of both axes is noticeable and the small deviation (non-zero intercept) is probably due to the experimental uncertainty in data acquisition and/or not uncontrolled imperfections. The same characteristics are observed for all five pumping wavelengths.

For each pump wavelength, we calculate the slopes $m$ of the linear regression curves between pump $S_3$ and PL $S_3$ similar to the one shown in Figure 2 (b). The slopes represent also the value of the maximally achievable spin polarization as $S_3(\pm1) =$

± *m*. We expect this to be a more robust estimator of spin polarization than taking the values for maximal pump helicity alone as it includes information from the whole characteristic. These obtained maximally achievable spin polarizations are shown in Figure 3 as a function of pump wavelength. Data from different runs at each wavelength are clustered together indicating the reliability of the results. Secondly and importantly, there is clear variation in the achievable spin polarization when changing the pump wavelength. Note that this is the spin polarization of the PL and the one relevant to implement a spin laser. The spin polarization in the QD ground state will be higher as some mixing between heavy-hole (HH) and light-hole (LH) states and admixture between spin-up and spin-down states due to spin-orbit coupling will lead to depolarization of the PL [24, 34, 55].

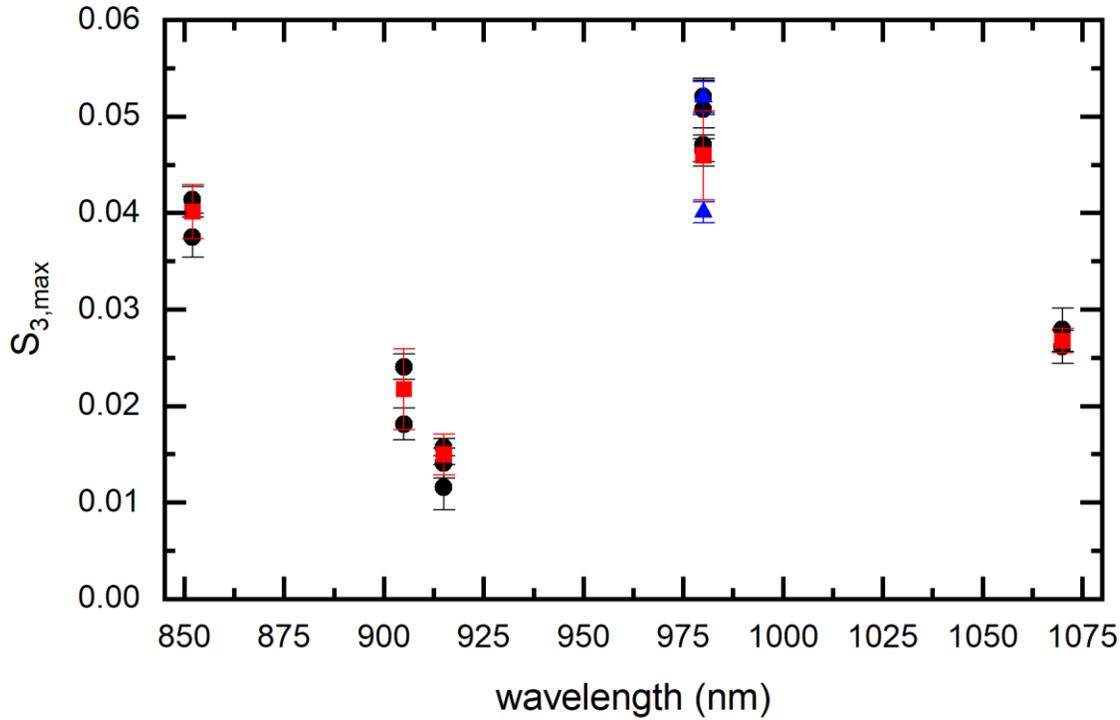

FIG. 3 Maximally achievable spin polarization of PL as obtained from the slope of PL $S_3$ vs pump $S_3$ (black circles). The error bars stem from the uncertainty returned by the linear regression fitting routine (Microcal OriginPro 2022) using instrumental weighting (i.e. proportional to inverse variance) for the individual data points. The red squares are a weighted average over all data at this wavelength. The black data are obtained in the same area of the wafer, the two triangular blue data points at 980 nm are probing two other locations on the same wafer. For the best spin polarization for 980 nm pumping further data were taken at different positions on the sample, respectively another piece of the wafer. Although variations between different run are larger than the statistical uncertainty of a single run, the clustering of the data indicate robust results. See [supplementary material], Section III for details of data analysis.

We estimate an effective spin polarization lifetime ($\tau_s$) from the slope, respectively from $S_{3,\text{max}}$, via the following relation [56, supplementary material]:

$$m = \frac{1}{1+\frac{\tau}{\tau_s}}, \tag{3}$$

where *m* is the slope of $S_3$ dependence shown in Figure 3 and $\tau$ is the carrier lifetime taken to be 770 ps from measurements on similar samples [57]. This relation follows from the spin-flip model taking only the heavy-hole dynamics into account, i.e. assume perfect optical selection rules [15]. Note that $\tau_s$ as introduced in [15] describes the relaxation in the ground state alone. Here we use it to describe an effective lifetime as also the spin relaxation occurring during the energy relaxation from the pumping energy to the QD ground state is lumped into it. This approach was used before in [7, 56] and provides a lower limit to the relaxation in the QD ground state alone. It will provide some indication of the performance of a spin laser as it gives an indication of the level of spin polarization achievable in the ground state for a given spin polarization of the external pump. In general, we caution that the spin lifetime within the ground state and the pump spin polarization reaching the ground state are different parameters and might influence the dynamics of a spin laser in different ways. If the intrinsic spin polarization in the QD ground state is known by time domain measurements, the deviation between the intrinsic and effective lifetime will give an indication of the loss of spin coherence during energy relaxation. The calculated effective spin polarization lifetimes as a function of the pump wavelength are shown in Figure 4.

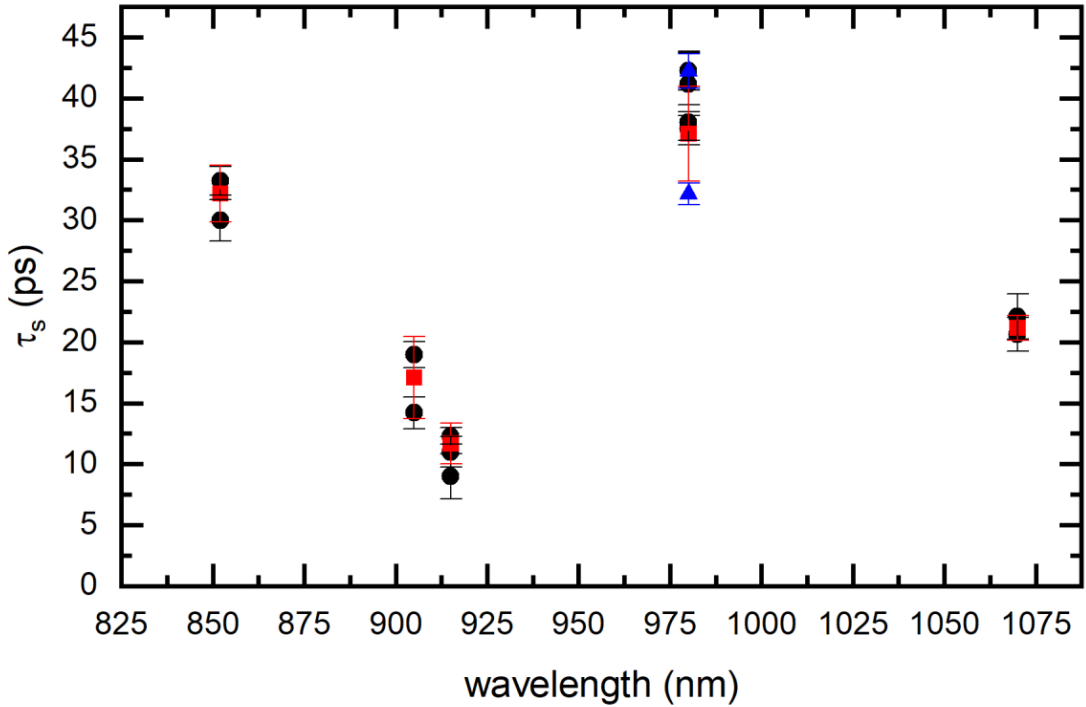

FIG. 4. Calculated effective spin polarization lifetime as a function of pump wavelength. Black/blue dots: from individual measurement, red squares: weighted average for each wavelength (as in Fig. 3).

Although there is some scatter of individual measurements, it is evident that there is a strong dependence of the achievable spin polarization, respectively effective polarization lifetime, on the pump wavelength. The highest spin polarization of 0.046±0.005 or effective lifetime of 38±4 ps is obtained for pumping at 980 nm. This can be expected as this is at the low energy edge of the InGaAs QW and hence absorbing states will have dominantly HH character [24]. Spin polarization is a factor of 2.3 to 3.4 lower at 905-915 nm. This is probably due to the admixture of LH states with the opposite spin selection rules as the HH [24]. From the relative strength of the HH-LH transition of 3:1 one expects a reduction of spin polarization after absorption by a factor of 2 [24]. Reports in the literature vary. For p-doped InAs/GaAs QD, Ref. [56] also reports an enhancement of spin polarization pumping the low energy vs the high energy wetting layer states, whereas Refs. [25, 37] do not find a significant difference. Ref. [37] relates this to thickness fluctuations mixing HH and LH states. In our case, we are pumping into a DWELL and not a wetting layer (as in [35,53]), i.e. a thicker structure less susceptible to interface fluctuations. For spin lasers, achieving a high spin polarization is an obviously important figure of merit but other factors like achieving an overall efficient pumping and avoiding excess heat are also important for laser design. For the laser gain structures considered here, the overall quantum efficiency is a factor 2-3 higher for pumping at 915 nm than at 980 nm whereas the spin polarization is a factor of 2-3 lower, i.e. a trade-off between optimal overall pumping efficiency and optimal spin polarization, which needs to be optimized for the specific spintronics application one is targeting.

Excitation at 1070 nm yields a spin polarization of 0.027±0.001 (effective lifetime of 21±1 ps), in between the values for 980 nm and 905-915 nm. Absorption will be either in very high-level QD states or disorder induced states [58, 59] of the QW, both being expected to have strong HH characteristics. However, it should be noted that the excitation is very inefficient (only about $2\times10^{-3}$ of best value at 915 nm). The spin polarization for pumping at 852 nm is quite high, 0.040±0.003 (effective lifetime of 32±3 ps). This is surprising and merits further investigation, as the main absorbing states are GaAs bulk states where LH and HH are degenerate and one expects a comparatively low efficiency. Indeed, a drop from 35% spin polarization for excitation in the HH WL states to 5% spin polarization for barrier excitation was observed in p-doped samples in [56]. The highest measured effective lifetimes incorporating relaxation during energy relaxation into the QD and relaxation within the QD is 40 ps and about a factor of 2 shorter than the lifetime in the QD ground state alone of about 70-80 ps [26, 37, 49] (up to 120 ps [38]) for InAs/Gas (not DWELL) QD.

In conclusion, the investigations confirm a fairly high CW spin polarization for DWELL vertical cavity spin structures making them suitable for macroscopic and miniature spin-VECSEL. For the latter, they provide the confirmation of material parameters enabling the robust spin amplification already observed above threshold [7]. As highest spin polarization (at 980 nm) is not achieved for highest pump efficiency (at 915 nm), the optimal wavelength for pumping the quantum well is shown to depend on whether low threshold and low power consumption or robust spin polarization are the more important figures of merit for a particular application. Pumping the barriers at 852 nm or states between QW and QD ground state at 1070 nm yielded surprisingly high spin polarization and it appears to be fruitful to look into mechanisms further, but they probably do not provide pump options for good lasing devices.

## SUPPLEMENTARY MATERIAL

Supplementary material provides detailed information on the quantum dot VECSEL sample structure, including layer composition and photoluminescence characterization. It also contains a description of the connection between spin polarization and spin lifetime, detailed data analysis procedures, and additional supporting figures for the polarization-resolved measurements.

## ACKNOWLEDGMENTS

We are grateful to Supakorn Phutthaprasartporn, Antonio Hurtado and Simon Armstrong for early contributions to the development of the experiment and to Mike Adams and Ian Henning for making the samples available. TA is grateful to Arnaud Garnache for useful discussion on spin relaxation in VECSEL structures. JD has been supported by European Training Network ColOpt, which was funded by the European Union (EU) Horizon 2020 programme under the Marie Skodowska-Curie action, grant agreement 721465. SA also extend his appreciation to Taif University, Saudi Arabia, for supporting this work through project number (TU-DSPP- 2024-159). We are grateful to Jonathan Pritchard for providing access to the 852 and 1070 nm pump sources.

## CONFLICT OF INTEREST

The authors have no conflicts to disclose